\documentclass[aps,prl,preprint,superscriptaddress]{revtex4}
\usepackage{graphicx}
\usepackage{color}

\begin{document}

\title{Direct Measurement of Surface Transport on a Bulk Topological Insulator}
\author{Lucas Barreto}
\affiliation{Department of Physics and Astronomy, Interdisciplinary Nanoscience Center (iNANO), Aarhus University,
8000 Aarhus C, Denmark}
\author{Lisa K\"uhnemund}
\author{Frederik Edler} 
\author{Christoph Tegenkamp}
\affiliation{Institut f\"ur Festk\"orperphysik, Leibniz Universit\"at Hannover, 30167 Hannover, Germany}
\author{Jianli Mi}
\author{Martin Bremholm}
\author{Bo Brummerstedt Iversen}
\affiliation{Center for Materials Crystallography, Department of Chemistry, Interdisciplinary Nanoscience Center (iNANO), Aarhus University,
8000 Aarhus C, Denmark}
\author{Christian Frydendahl}
\author{Marco Bianchi}
\affiliation{Department of Physics and Astronomy, Interdisciplinary Nanoscience Center (iNANO), Aarhus University,
8000 Aarhus C, Denmark}
\author{Philip Hofmann}
\affiliation{Department of Physics and Astronomy, Interdisciplinary Nanoscience Center (iNANO), Aarhus University,
8000 Aarhus C, Denmark}
\affiliation{email: philip@phys.au.dk}

\begin{abstract}
Topological insulators are guaranteed to support metallic surface states on an insulating bulk, and one should thus expect that the electronic transport in these materials is dominated by the surfaces states. Alas, due to the high remaining bulk conductivity, surface contributions to transport have so-far only been singled out indirectly via quantum oscillations \cite{Qu:2010,Analytis:2010b}, or for devices based on gated and doped topological insulator thin films, a situation in which the surface carrier mobility could be limited by defect and interface scattering \cite{Steinberg:2010,Checkelsky:2011,Kim:2012,Xia:2013}. Here we present the first direct measurement of surface-dominated conduction on an atomically clean surface of bulk-insulating Bi$_2$Te$_2$Se. Using nano-scale four point setups with variable contact distance, we show that the transport at 30~K is two-dimensional rather than three-dimensional and by combining these measurements with angle-resolved photoemission results from the same crystals, we find a surface state mobility of 390(30)~cm$^{2}$V$^{-1}$s$^{-1}$ at 30~K at a carrier concentration of 8.71(7)$\times 10^{12}$~cm$^{-2}$.   
\end{abstract}

\date{\today}

\maketitle

The most significant hurdle for reaching surface-dominated transport in a topological insulator (TI) is currently believed to be the high conductivity of the underlying bulk. In fact, most bulk TIs have a small band gap, are degenerately $n$-doped, and/or suffer from poor screening of impurities \cite{Skinner:2012}. Only recently, materials  with a semiconducting temperature dependence of the bulk resistivity $d\rho / dT$ have been synthesised \cite{Ren:2010,Mi:2013}. While the high bulk conductance is a well-recognised issue, it is by no means obvious that the surface conductance would be very high. For a one-dimensional (1D) edge state of a two-dimensional (2D) TI (the quantum spin Hall effect), the spin texture provides a protection from back scattering and since this is the only possible scattering process in 1D, perfect quantum transport results at low temperature \cite{Bernevig:2006,Konig:2007}. For a 2D surface state, backscattering is still forbidden but other scattering processes are not, such that transport could be strongly affected by impurity scattering and, at finite temperature, also by electron-electron and electron-phonon scattering. Indeed, the heavy elements in the topological insulators typically lead to a low Debye temperature, increasing the significance of phonon scattering \cite{Hatch:2011}. Achieving the regime of surface-dominated conductance and quantifying the  surface mobility is therefore an important goal of the field. 

\begin{figure}
 \includegraphics[width=.75\textwidth]{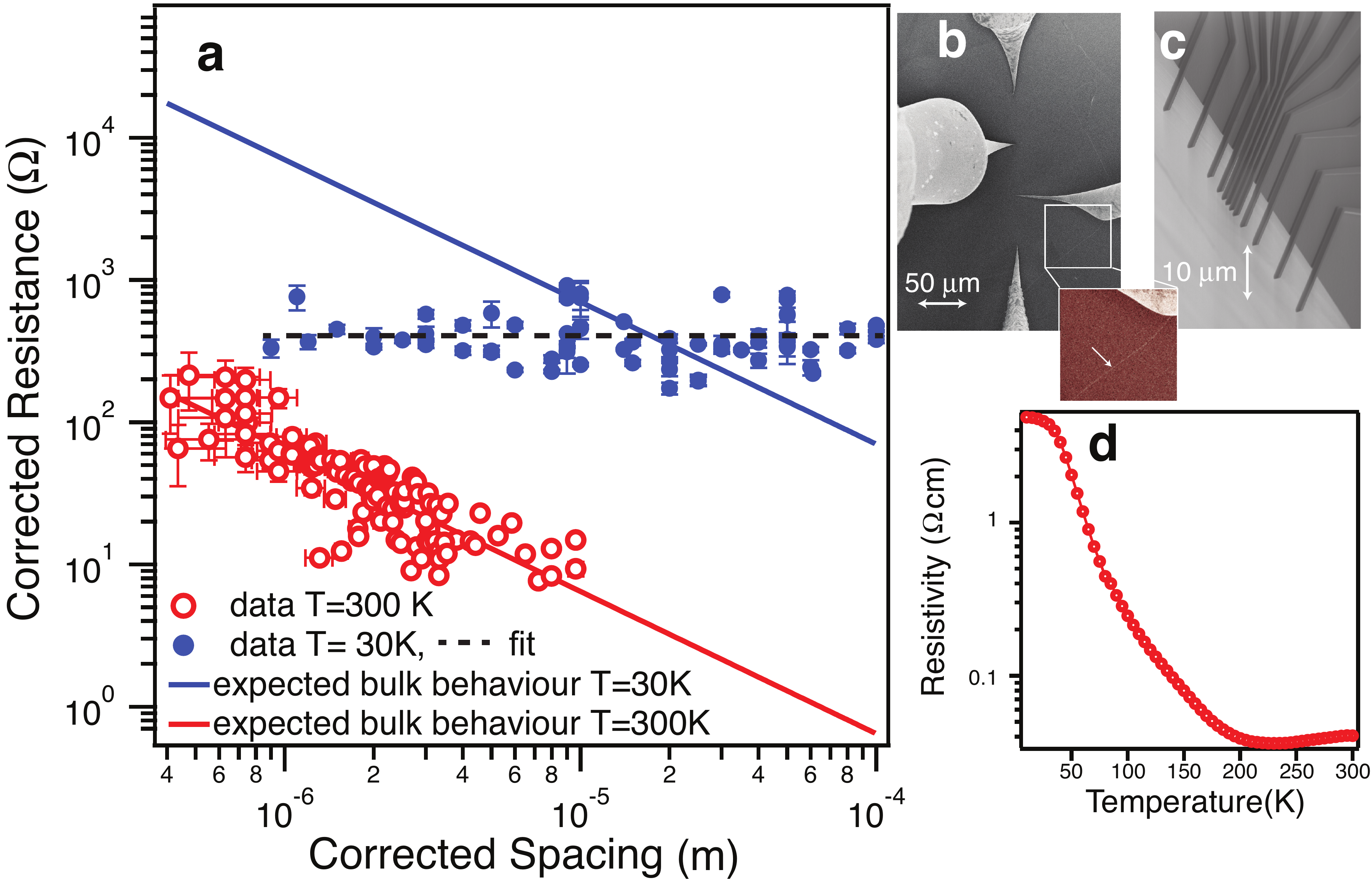}\\
  \caption{(a) Four point resistance measurements on a clean Bi$_2$Te$_2$Se surface taken at 300~K and at 30~K (data points), together with the expected result for bulk dominated transport at these two temperatures (solid lines). Note that the statistical uncertainties are in some cases smaller than the markers. The dimensionality of the transport can be read directly from the data as 2D for 30~K and 3D for 300~K. (b) Scanning electron microscopy (SEM) image of the four STM contacts used for the 30~K measurements on the surface of Bi$_2$Te$_2$Se. The inset shows how surface steps appear in the image (marked by an arrow). (c) SEM image of the twelve point probe used for the 300~K measurements. (d)  Bulk resistivity of the Bi$_2$Te$_2$Se sample as a function of temperature.    }
  \label{fig:1}
\end{figure}

In the present work, the direct measurement of surface transport is achieved by combining high-quality bulk-insulating crystals of  Bi$_2$Te$_2$Se \cite{Mi:2013} with four point transport measurements on the clean surface in ultrahigh vacuum and on a variable length scale. The transport measurements shown in Fig. \ref{fig:1}(a) were performed by two instruments: a four tip scanning tunnelling microscope (4STM) \cite{Shiraki:2001,Guise:2005,Baringhaus:2013} and a collinear twelve-point probe (12pp) \cite{Gammelgaard:2008,Perkins:2013}. These are shown in Fig. \ref{fig:1}(b,c). For the 4STM, the scanning electron microscopy image is that of the four contacts placed on the actual surface of  Bi$_2$Te$_2$Se. Typically, transport measurements were performed on large areas free of atomic steps (with lateral dimensions $\gtrsim 100 \times 100$~$\mu$m$^2$) but this particular image has been chosen to show how such steps appear in the images (see magnified inset). 

Fig. \ref{fig:1}(a) shows the key results of this paper, the corrected four point resistance $R$ as a function of the corrected contact spacing $s$ at 300~K and at 30~K (see Methods section for a detailed discussion of the corrected resistance and corrected contact spacing). The data points at 300~K have been collected with the 12pp whereas the 30~K data points have been measured with the 4STM. At 300~K, the four point resistance $R$ is observed to be inversely proportional to the contact distance $s$. This behaviour is characteristic  for four point measurements measurements on a 3D bulk crystal, where we expect that $R=R_{3D}=1/(2\pi s \sigma_b)$, with $\sigma_b$ being the bulk conductivity \cite{Hofmann:2009}. The red line through the data points shows the expected four point probe resistance  at room temperature, based on a macroscopic bulk resistivity measurement on the same crystal (see Fig. \ref{fig:1}(d)). Clearly, the agreement is excellent and we find no indication of surface transport. 

The situation is entirely different for the measurements taken at 30~K. Here the measured $R$ is approximately independent of $s$. Such a behaviour is expected for pure 2D conductance where one finds that $R=R_{2D}=\ln 2 /( \pi \sigma_s)$, with $\sigma_s$ being the 2D the sheet conductance \cite{Hofmann:2009}.  If we assume the total four point resistance to be the contribution of surface and bulk in parallel, i.e. $R_{tot}^{-1}= R_{2D}^{-1}+R_{3D}^{-1}$, it thus appears that bulk transport contribution is negligible at this temperature. When taking the average resistance at 30~K (dashed black line in Fig. \ref{fig:1}(a)), we obtain a sheet conductance of $\sigma_s=5.3(3)\times10^{-4}$~$\Omega^{-1}$.

Again, we can calculate the expected behaviour for bulk-dominated transport. From Fig. \ref{fig:1}(d) we find the bulk resistivity at 30~K to be $\approx 4~\Omega$cm, consistent with the highest values reported in the literature \cite{Ren:2010}, and two orders of magnitude higher than at 300~K. Using this value, a calculation of the expected bulk behaviour results in the blue line. The four point resistance can clearly not be reconciled with bulk behaviour, not only because of the wrong dependence on the contact spacing, but also because the surface is much more conductive at small contact spacings than expected for the bulk solid. Interestingly, the blue line crosses the constant of the dashed line through the data points for a spacing of $\approx 20$~$\mu$m. One should therefore expect a transition to bulk-dominated behaviour around this spacing but such a trend is not observed. The surface-dominated transport is seen to persist to the largest measured spacing of  $\approx 100$~$\mu$m, suggesting that the actual bulk resistivity of the underlying sample \emph{in the probed surface area} is larger than the value obtained from the macroscopic resistivity measurement. In fact, it is quite likely that a macroscopic measurement for a not perfectly homogeneous sample \cite{Mi:2013} will tend to give an average resistivity that is dominated by percolated regions of low resistance. Moreover, the surface region for the microscopic measurements was deliberately chosen to avoid areas with steps, impurity clusters or other irregularities. Since the data shows no sign of a deviation from the constant resistance even at the largest spacings ($\approx 100$~$\mu$m), we conclude that the local bulk resistivity is at least one order of magnitude higher than that obtained from the bulk measurement. 

\begin{figure}
\includegraphics[width=.75\textwidth]{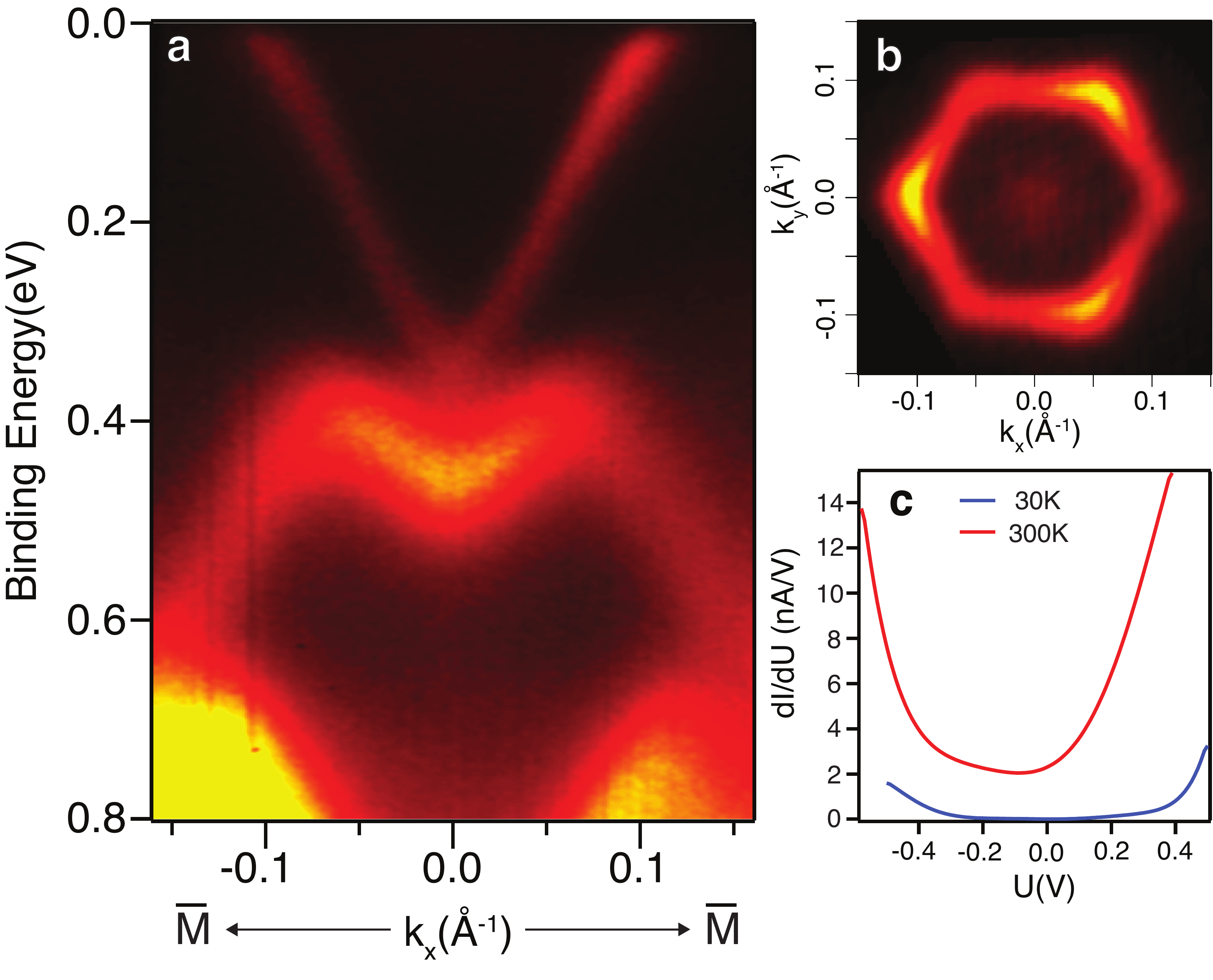}\\
  \caption{ARPES and scanning tunnelling spectroscopy characterisation of the Bi$_2$Te$_2$Se samples used in this study. (a) Surface state dispersion along the $\bar{M}-\bar{\Gamma}-\bar{M}$ direction.  (b) Photoemission intensity at the Fermi energy. (c) Scanning tunnelling spectra taken at 300~K and at 30~K.} 
  \label{fig:2}
\end{figure}

In order to determine the surface state mobility from the sheet resistance, it is necessary to know the carrier concentration in the surface state. To this end, the surface electronic structure of the  Bi$_2$Te$_2$Se is determined by  angle-resolved photoemission spectroscopy (ARPES). Fig. \ref{fig:2}(a) and (b) show the results of such a measurement. The spectra clearly show the existence of the topological surface state with the Dirac point at a binding energy of approximately $360$~meV and a considerable hexagonal warping of the constant energy contour at the Fermi energy. The distance from the hexagon's edges to the centre is found to be $0.0997(4)$~\AA$^{-1}$, corresponding to a density of $8.71(7)\times10^{12}$ electrons per cm$^2$. From this and $\sigma_s$ determined above, we can evaluate the carrier mobility to be  390(30)~cm$^{2}$V$^{-1}$s$^{-1}$. This surface mobility is somewhat lower (by a factor of $\approx 3$) than the low temperature values reported for  Bi$_2$Se$_3$ \cite{Kim:2012} or Bi$_2$Te$_2$Se \cite{Ren:2010}. Apart from the obvious reason of increased electron-phonon scattering at the higher temperature here (30~K instead of 1.6~K for the Bi$_2$Te$_2$Se results in Ref. \cite{Ren:2010}), the lower mobility could be caused by the higher carrier density ($8.71(7)\times10^{12}$ instead of $1.5\times10^{12}$ in Ref. \cite{Ren:2010}) and thereby increased phase space for scattering. 

When interpreting $\sigma_s$ as solely originating from the topological surface state, it is important to exclude the contributions of two-dimensional electron gases (2DEGs) that can potentially form at the surfaces of  TIs due to adsorbate-induced band bending \cite{Bianchi:2010b,Bianchi:2011}. ARPES can identify both degenerately populated conduction band states near the surface as well as 2DEGs. Both give rise to features at the Fermi energy within the Fermi contour of the topological surface state. Bulk conduction band states appear as a broad feature that disperses with photon energy, while 2DEGs have a well-defined contour that does not show such dispersion. The data in Figs. \ref{fig:2}(a) and (b) have been taken with a photon energy that maximises the sensitivity towards bulk states and 2DEGs and we can exclude the presence of a 2DEG. Only a very weak and diffuse enhancement of the photoemission intensity is seen around normal emission ($k_{\parallel}=0$~\AA$^{-1}$) in Fig. \ref{fig:2}(b) (but not in Fig. \ref{fig:2}(a) because the colour scaling of this figure is dominated by the intense states at higher binding energy). After several hours in vacuum, however, a small amount of adsorbate-induced band bending can be seen, leading to the observation of the bulk conduction band minimum at $k_{\parallel}=0$~\AA$^{-1}$ but not to a distinct 2DEG.  In fact, we find Bi$_2$Te$_2$Se surfaces to be much less sensitive to contamination-induced band bending than Bi$_2$Se$_3$ surfaces and no 2DEG formation is observed, unless the surface is doped with alkali atoms. 

The 2D character of the four point resistance $R$ at 30~K appears to rule out a significant contribution to the conductance from bulk states and we can support this by further observations. Fig. \ref{fig:2}(c) shows  scanning tunnelling spectra, taken with one of the tips of the 4STM at 30~K and 300~K, around 4~hours after cleaving the sample. While distinct signatures of the topological states are typically difficult to identify in such spectra, similar to the situation in Ref. \cite{Nurmamat:2013}, it is clear that the Fermi energy at the surface is placed well within the bulk band gap and the bulk states are therefore not expected to contribute significantly to the observed transport. Moreover, the in-gap differential conductance at 30~K is much lower than at 300~K, consistent with the resistivity change over this temperature range.

A  time-dependent near-surface band bending can be expected to increase the bulk carrier density near the surface and hence the conductance. However, we could not observe any systematic time-dependence of the measured $\sigma_{2D}$ within the first 30~hours after the cleave. Indeed, no such change could be observed even when intentionally contaminating the surface with carbon monoxide at a partial pressure of $\approx 5 \times 10^{-9}$~mbar for two hours. From this we conclude that while a band bending might be present, the contribution of bulk states to the transport is still negligible. There are two factors explaining this: The band bending-induced carrier density is substantially smaller than that from the topological state and it is likely that the bulk states suffer from stronger scattering than the surface states. 

In conclusion, we have reported a direct surface-state dominated transport measurement on the clean surface of bulk-insulating Bi$_2$Te$_2$Se. The surface mobility is found to be high, 390(30)~cm$^{2}$V$^{-1}$s$^{-1}$, at the measurement temperature of 30~K. Reaching the surface-dominated transport regime has been possible by the combination of a nano-scale four point probe technique with high-quality bulk crystals. The results indicate that the bulk resistivity in a carefully chosen region of the crystal may significantly exceed the value measured from a macroscopic bulk crystal. This is not surprising in view of the fluctuations of physical properties observed in large bulk crystals \cite{Mi:2013} and it suggests that pushing surface-dominated transport towards room temperature could be an achievable goal. 

\section{Methods}

Bi$_2$Te$_2$Se crystals were grown and characterised as described in Ref. \cite{Mi:2013}. Bulk transport measurements were performed on mm-size crystals using a Quantum Design Physical
Property Measurement System. For the surface transport measurements, the samples were cleaved in the ultra-high vacuum chamber. Nano-scale transport measurements with the 4STM were carried out using the four STM tips in a collinear and equidistant probe configuration (probe spacing $s$) with the current passed through the outer two probes and the voltage measured over the inner two. The position of the tips on the surface and the surface morphology was controlled using a scanning electron microscope. The instrument is described in detail in Ref. \cite{Guise:2005}. The 12pp measurements, on the other hand, were taken with a monolithic collinear probe with different contact distances \cite{Gammelgaard:2008}.  Using such a probe, it is not possible to vary the contact spacing but it is possible to effectively emulate an equidistant four point probe measurement. To this end, one introduces the corrected resistance $\chi_{2D} R_{total}^{4pp}$ and the corrected spacing $s_{3D}^{eq.}/\chi_{2D}$. They correspond to the measured resistance $R$ and contact spacing $s$ for an equidistant four point probe where the outer two contacts are used as current sources and the inner two as voltage probes, i.e. to the situation in the 4STM measurements. This approach has been used to display the 12pp data points in Fig. \ref{fig:1}, whereas the 4STM data points in the figure are merely the uncorrected four point resistance and contact spacing. For a detailed treatment of the concepts of corrected resistance and corrected spacing, see Ref.  \cite{Wells:2008c,Perkins:2013}. The vertical error bars on the points represent the uncertainty in the resistivity measurement and the horizontal error bars the uncertainty that arises from an in-line misplacement of the contacts. Note that the scatter in the data points is higher than what would be expected from the error bars. This is a well known effect and can be ascribed to sample inhomogeneity \cite{Perkins:2013}.

ARPES experiments were performed at the SGM-3 beamline of the synchrotron radiation source ASTRID \cite{Hoffmann:2004}.  The sample temperature during the ARPES measurements was 70~K. The energy and angular resolution were better than 20~meV and 0.2$^{\circ}$, respectively. The photon energy (17.5~eV) was chosen such that the photoemission intensity from the bulk conduction band minimum is maximised. The carrier density in the topological surface state was estimated from the size of the measured Fermi surface.

\section{Acknowledgements}
We gratefully acknowledge support from the Carlsberg foundation, the VILLUM foundation, the Danish National Research Foundation and Capres A/S.




\end{document}